\documentclass[aps,prb,twocolumn,showpacs,showkeys,amssymb]{revtex4}

\usepackage{graphics}
\usepackage{bm}
\usepackage{amsmath}
\usepackage{subfigure}

\newcommand{\wt}{\widetilde}
\newcommand{\be}{\begin{equation}}
\newcommand{\ee}{\end{equation}}
\newcommand{\bea}{\begin{eqnarray}}
\newcommand{\eea}{\end{eqnarray}}
\newcommand{\vect}[1]{\mathbf{#1}}

\begin{document}
\title{Effect of coexisting order of various form and wave vector on low-temperature thermal conductivity in $d$-wave superconductors}

\author{Philip R. Schiff}
\author{Adam C. Durst}
   \affiliation{Stony Brook University, Stony Brook, NY 11794-3800, USA}
    \email{pschiff@grad.physics.sunysb.edu}

\begin{abstract}
In light of recent experimental evidence of density wave order in the cuprates, we consider a phenomenological model of a $d$-wave superconductor with coexisting charge, spin or pair density wave order of various form and wave vector. We study the evolution of the nodal structure of the quasiparticle energy spectrum as a function of the amplitude of the coexisting order and perform diagrammatic linear response calculations of the low-temperature (universal-limit) thermal conductivity. The work described herein expands upon our past studies, which focused on a particular unit-cell-doubling charge density  wave, generalizing our techniques to a wider class of coexisting order. We find that the question of whether the nodes of the $d$-wave superconductor survive amidst a reasonable level of coexisting order is sensitive to the form and wave vector of the order. However, in cases where the nodes do become gapped, we identify a signature of the approach to this nodal transition, in the low-temperature thermal conductivity, that appears to be quite general. The amplitude of this signature is found to be disorder-dependent, which suggests a connection between the presence of coexisting order in the underdoped cuprates and recent observations of deviations from universal (disorder-independent) thermal conductivity in the underdoped regime.
\end{abstract}

\maketitle
\section{Introduction}
The low energy excitations of cuprate superconductors are Dirac fermions, which arise due to the $d$-wave nature of the superconducting order parameter. One expected signature of these nodal quasiparticles is the presence of a universal term in the low-temperature thermal conductivity, $\kappa_{00}$, which depends only on the ratio of the gradient of quasiparticle dispersion to the gradient of the gap, $v_f/v_\Delta$, but not on the disorder\cite{lee01,dur01,gra01}. In the optimally doped and overdoped regimes, $\kappa_{00}$ has been observed in many instances, and agrees closely with its predicted value\cite{tai01,chi01,chi02,nak01,pro01,sut01,hil01,sun01,sut02,haw01,sun02}.
For instance, in $\mathrm{YBa_2(Cu_{1-x}Zn_x)_3O_{6.9}}$, $\kappa_{00}$ is observed to be insensitive to the concentration of Zn atoms, which are varied to allow the scattering rate to range over several orders of magnitude\cite{tai01}. However, in some cases,  the value of the universal limit thermal conductivity, $\kappa_{00}$, is different than expected, or not observed at all, in particular as one approaches the underdoped regime\cite{hus01,haw02,and01,sun03,sun04,pro02}.
One possible reason may be that disorder-induced local magnetic moments enhance the scattering rate while leaving the density of states unaffected, thus reducing the transport\cite{ander01}.

Another mechanism which might account for deviations from the expected value of $\kappa_{00}$ is the presence of competing order parameters. For years, evidence of the presence of additional symmetry breaking order parameters in cuprates has been compiled in neutron scattering data and scanning tunneling microscopy experiments\cite{hof01,hof02,how01,ver01,mce01,han01,mis01,mce02,koh01,boy01,han02,pas01,wis01,koh02,kiv01,fis01}. The presence of additional orders other than superconducting may be incidental, yet it also may be intrinsically related to the complex phenomenon of high temperature superconductivity itself\cite{cha01}.

The addition of order parameters which preserve time-reversal symmetry followed by a lattice translation was found to preserve the nodal structure of the quasiparticles, for small amplitudes of order\cite{ber01}. As the strength of such ordering perturbations increases, the locations of the nodal excitations evolve in $k$-space. For sufficiently large amplitude, the quasiparticle spectrum can be entirely gapped\cite{par01,ber01,dur02,sch01}. Such a modification of the quasiparticle spectrum should manifest itself in the low temperature thermal conductivity\cite{dur02,sch01,gus01}.

In this paper, we model a cuprate superconductor using a mean-field formalism describing a BCS-like $d$-wave superconductor ($d$SC) perturbed by an additional order parameter. We calculate the low-temperature thermal conductivity, accounting for the presence of several different varieties of
competing orders. We argue that these predictions can then be used as an indirect verification of the presence or absence of various competing orders in cuprates.

 A previous linear response calculation of $\kappa_{00}$ in a $d$SC with the addition of a $\vect{Q}=(\pi,0)$ charge density wave showed that vertex corrections were not important for the universal limit thermal conductivitity, within the self-consistent Born approximation\cite{sch01}. As the charge density wave's amplitude increased beyond a critical value $\psi_c$, the quasiparticle spectrum became gapped. Correspondingly, the thermal conductivity (made anisotropic by the presence of the density wave) vanished beyond that critical strength of ordering. In addition, a dependence on disorder resulted, in particular for charge orderings near the transition. Armed with this information, we proceed to study the effects of a wider class of density waves on the  low-energy properties of cuprates.

In Sec.~\ref{sec4:model}, we will develop the mean-field formalism we will use to describe superconductors in the presence of competing orders. We write effective hamiltonians for charge, spin and pair density waves of several wave-vectors. Additionally, we describe configurations with multiple wave vectors, such as checkerboard order. Next, in Sec.~\ref{sec:thermalcond} we will derive the current operators associated with the various kinds of orders, and use this to establish a relation for the bare-bubble thermal conductivity. Finally in Sec.~\ref{sec:results}, we will apply our results to several different cases, and compare and contrast the results.
\section{Model}
\label{sec4:model}
\subsection{States of broken symmetry}
States that arise as a result of broken symmetries are characterized by the presence of non-vanishing off-diagonal matrix elements. The superconducting state itself, for instance, can be identified with the non-vanishing anomalous Green functions, as was shown by Gor'kov\cite{schrieffer01}. These anomalous Green functions are defined in space and time as
\be
<\psi_\alpha(r,t)\psi_\beta(0,0)>.
\ee
Given singlet paired electrons of opposite momenta, this corresponds, in momentum space, to
\be
<\psi_\alpha(k,t)\psi_\beta(-k,t)>.
\ee
In a similar fashion, ordered states representing density waves will also admit non-vanishing correlations between states separated by the wave-vector of the
density wave. In this chapter, we will consider the subset of those which are defined in momentum space as
\be
<\psi^{\alpha\dagger}(k+Q,t)\psi_\beta(k,t)>\equiv\Phi_Q f(k) d(\alpha,\beta),
\ee
representing charge $(d=\delta^\alpha_\beta)$ and spin ($d=\delta^\alpha_\beta(\delta^{\alpha}_\uparrow-\delta^\alpha_\downarrow)$) density waves, as well as
pair density waves
\be
<\psi^{\alpha\dagger}(k+Q,t)\psi^\dagger_\beta(k,t)>\equiv\Phi_Q f(k)\epsilon^\alpha_\beta.
\ee
For the purposes of simpler classification of orders, we are carefully following some definitions made by Nayak in Ref.~\onlinecite{nay01}, so that $\Phi_Q$ will represent the magnitude and phase of the density wave and $f(k)$ is an element of a representation of the space group of $\vec{Q}$ on a square lattice.
Certain order parameters obey restrictions. For instance, charge and spin density waves for which $2\vect{Q}$ is a member of the reciprocal lattice obey the
additional condition
\be
\label{eq:commensurate}
\Phi_Q f(k+Q)=\Phi^*_Q f^*(k)
\ee
as was pointed out in Ref.~\onlinecite{nay01}.

Written as a sum over real space, the hamiltonian representing a charge density wave system is
\be
H_{CDW}=\sum_{\substack{r r\prime\\\sigma}}\psi e^{-i\vec{Q}\cdot(\vec{r}-\vec{r}_0)}f(r-r\prime)c_{r\sigma}^\dagger c_{r\prime\sigma} + \mathrm{h.c.}.
\ee
Upon Fourier transform this becomes
\be
H_{CDW}=\sum_{k\sigma}(\Phi_Q f_k c_{k+Q\sigma}^\dagger c_{k\sigma}+\Phi_Q^* f_k^* c_{k\sigma}^\dagger c_{k+Q\sigma}),
\ee
with the definition $\Phi_Q=\psi e^{i\vec{Q}\cdot\vect{r}_0}$, where $\vec{r}_0$ describes the shift of the density wave from being site-centered
and $\psi$ is the amplitude of the density wave. Coupled with Eq.~(\ref{eq:commensurate}), this indicates restrictions on certain density waves'
registration with the lattice.
\subsection{$d$-wave superconductor}
Our starting point is a model for $d$-wave superconductors
\be
\label{eq:HdSC}
H=\sum_{k,\sigma}\Big( \epsilon_k c_{k\sigma}^\dagger c_{k\sigma} + \Delta_k c^\dagger_{k\uparrow}c^\dagger_{-k\downarrow}\Big) + \textrm{h.c.}
\ee
where the normal state dispersion is given by a tight-binding hamiltonian,
\be
\epsilon_k=-2t(\cos k_x+\cos k_y)-t^\prime \cos k_x\cos k_y -\mu.
\ee
and the superconducting order parameter is of $d_{x^2-y^2}$ symmetry,
\be
\Delta_k = \frac{\Delta_0}{2}(\cos(k_x)-\cos(k_y)).
\ee
As given, this hamiltonian has nodal excitations, which are located along the $d_{x^2-y^2}$ symmetry axis in the $(\pm \pi,\pm \pi)$ directions. The nodes'
 distance from the $(\pm \pi/2,\pm \pi/2)$ points is controlled by the chemical potential $\mu$. These quasiparticles are massless Dirac fermions in the sense that they have conical dispersion. The excitation energy is
  \be
  E_k=\sqrt{\epsilon_k^2+\Delta_k^2},
  \ee
and at low energies, $\epsilon_k\sim v_f k_1$ and $\Delta_k \sim v_\Delta k_2$, where $k_1$ and $k_2$ are momentum-space displacements from the node in directions perpendicular and parallel to the Fermi surface respectively, $v_f$ is the Fermi velocity, and $v_\Delta$ is the slope of $\Delta_k$ at the node. For $\mu$ on the order of $t$ or smaller, the ratio of Fermi velocity to gap velocity is given as
\be
\frac{v_f}{v_\Delta}\approx\frac{4\sqrt{t^2-\frac{\mu}{t}t^{\prime^2}}}{\Delta_0}.
\ee
 Then, as perturbations are turned on, the locations of the nodes evolve in $k$-space, while the stability of the nodes is generally preserved for
 non-nesting perturbations which preserve the composite symmetry of lattice translation followed by time-reversal, as was noted by Berg and Kivelson\cite{ber01}.
\subsection{Density waves of different wave vectors}
\begin{figure}[htbp]
  	\centerline{
  \resizebox{3.25 in}{!}
  {\includegraphics{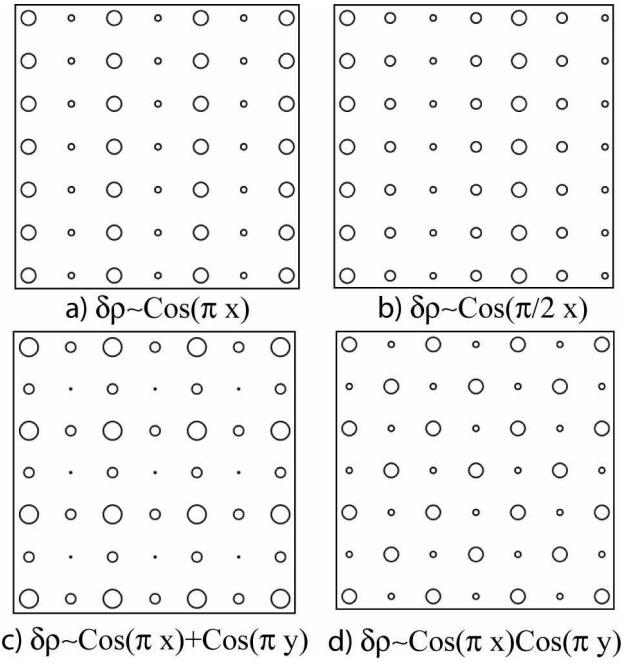}}}
  \caption[Position space illustration of four different density waves]{Illustrated are the four different density waves considered in this chapter, in real space. Each circle corresponds to the position of a Cu atom, and the size of the circle indicates whether the density at that site is higher or lower than the average. Illustrated are density waves of wave vector: (a) $\vect{Q}=(\pi,0)$, (b) $\vect{Q}=(\pi/2,0)$, (c) $\vect{Q}_1=(\pi,0)$, $\vect{Q}_2=(0,\pi)$ and (d) $\vect{Q}=(\pi,\pi)$}
  \label{fig:four_Qs_position}
\end{figure}

The presence of a uniform density wave in a superconductor changes the system in both real and momentum space. In real space, the unit cell increases. In momentum space, we see an effective reduction of the Brillouin zone, also called band folding. Accordingly, our second
quantized descriptions of the systems are modified. Whereas in a superconductor we can rewrite a quadratic effective hamiltonian using Nambu field operators,
\bea
\psi_k^\dagger=
\begin{pmatrix}
c_{k\uparrow}^\dagger & c_{-k\downarrow}
\end{pmatrix}
\hspace{30pt}
\psi_k =
\begin{pmatrix}
c_{k\uparrow} \\
c_{-k\downarrow}^\dagger
\end{pmatrix},
\eea
we can alternatively write extended Nambu vectors, such as
\bea
\psi_k^\dagger=
\begin{pmatrix}
c_{k\uparrow}^\dagger & c_{-k\downarrow} & c_{k+Q\uparrow}^\dagger & c_{-k-Q\downarrow}
\end{pmatrix}
\label{eq:extended_Nambu}
\eea
where the wave-vector $\vect{Q}$ connects each point in the first reduced Brillouin zone with a point in the second reduced Brillouin zone. Sums over $k$-space
are then performed by integrating over the reduced Brillouin zone (the shaded regions in Fig.\ref{fig:four_Qs_BZ}), and taking the trace of the now-extended
Nambu space matrix. The two descriptions are equivalent, but the extended Nambu description naturally fits the effective hamiltonians of systems with non-zero
mean-field density waves. In Fig.~\ref{fig:four_Qs_BZ}, we illustrate four different density waves which are considered in this paper:
$\vect{Q}=(\pi,0)$, $\vect{Q}=(\pi/2,0)$, $\vect{Q}=(\pi,\pi)$ and $\vect{Q}_1=(\pi,0)$, $\vect{Q}_2=(0,\pi)$ (checkerboard) orders. These disturbances are illustrated in real space in Fig.\ref{fig:four_Qs_position}.

\subsubsection{$\vect{Q}=(\pi,0)$ density waves}
\begin{figure}[htbp]
  	\centerline{
  \resizebox{3.25 in}{!}
  {\includegraphics{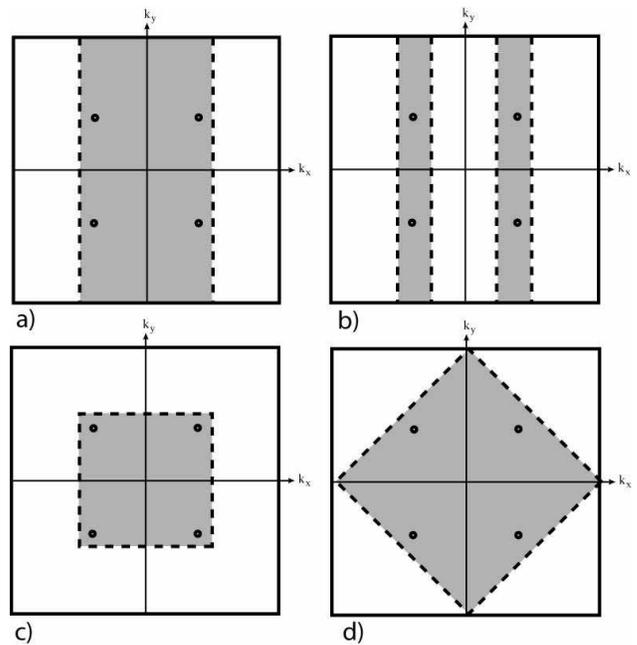}}}
  \caption[Brillouin zones of model $d$-wave superconductor with various density modulations]{Illustrated are the reduced Brillouin zones of a square lattice in the presence of different density waves. The dots illustrate approximately the location of nodal excitations in the original $d_{x^2-y^2}$-symmetry superconductor; the dashed line is the new zone boundary induced by the density wave. Illustrated are density waves of wave vector: (a) $\vect{Q}=(\pi,0)$, (b) $\vect{Q}=(\pi/2,0)$, (c) $\vect{Q}_1=(\pi,0)$, $\vect{Q}_2=(0,\pi)$ and (d) $\vect{Q}=(\pi,\pi)$}
  \label{fig:four_Qs_BZ}
\end{figure}

A density wave of wave vector $\vect{Q}=(\pi,0)$ corresponds to a striped system: the unit cell is doubled in the x-direction, and the Brillouin zone is reduced by
50$\%$ as seen in Fig.~\ref{fig:four_Qs_BZ}(a). The extended Nambu vector is that of Eq.~\ref{eq:extended_Nambu}, with $\vect{Q}=(\pi,0)$.
\subsubsection{$\vect{Q}=(\pi/2,0)$ density waves}
A density wave of wave vector $\vect{Q}=(\pi/2,0)$ corresponds again to a striped system, one in which the unit cell has increased by a factor of four,
and the Brillouin zone is reduced by the same factor. The reduced Brillouin zone is taken to be the region containing the pre-density-wave nodal
quasiparticle excitations of the $d$-wave superconductor; in Fig.~\ref{fig:four_Qs_BZ}(b) this is indicated with shading. The extended Nambu vector is
that of Eq.~\ref{eq:extended_Nambu_PIHALF}.
\subsubsection{$\vect{Q}_1=(\pi,0)$; $\vect{Q}_2=(0,\pi)$ checkerboard density waves}
Two density waves of equal weight in orthogonal directions corresponds to a checkerboard ordered system. As in the case of the $\vect{Q}=(\pi/2,0)$ case, the
Brillouin zone's area is reduced by a factor of four, although it is a different reduced Brillouin zone, illustrated in $k$-space in Fig.~\ref{fig:four_Qs_BZ}(c), and in real space in Fig.~\ref{fig:four_Qs_position}(c).
\subsubsection{$\vect{Q}=(\pi,\pi)$ density waves}
A density wave of wave vector $\vect{Q}=(\pi,\pi)$ corresponds to a system which is modulated in both the $k_x$ and $k_y$ directions: $\delta\rho\propto \sin(k_x)\sin(k_y)$. The reduced Brillouin zone is indicated in Fig.~\ref{fig:four_Qs_BZ}(d) as the shaded region, and the real space modulation is illustrated in Fig.\ref{fig:four_Qs_position}(d). The extended Nambu vector is as in Eq.~\ref{eq:extended_Nambu}, with $\vect{Q}$ now representing the $(\pi,\pi)$ density wave.

\subsection{Charge density waves}
A commensurate charge density wave is one for which the charge density is oscillatory in real space and repeats itself after translation by an integer number of lattice
constants. The momentum space description of a mean field hamiltonian for
such a system is
\be
\label{eq:HCDW}
H_{CDW}=\sum_{k\sigma}\Big(\Phi_Q f_k c_{k+Q\sigma}^\dagger c_{k\sigma} + \Phi_Q^* f_k^* c_{k\sigma}^\dagger c_{k+Q\sigma}\Big).
\ee
A charge density wave $\rho = \rho_0 + \delta \rho$ which doubles the unit cell (so that $\delta \rho $ alternates sign from cell to cell in the x-direction)
has wave-vector $\vect{Q}=(\pi,0)$. A $(\pi,0)$ CDW perturbation in its 4-component extended Nambu basis (particle, hole, shifted particle, shifted hole) is given by
\be
H_{CDW}=
\sum_k^{\prime}\psi_k^\dagger
\begin{pmatrix}
0 & 0 & A_k^* & 0 \\
0 & 0 & 0 & -A_{-k} \\
A_k & 0 & 0 & 0 \\
0 & -A_{-k}^* & 0 & 0
\end{pmatrix}
\psi_k,
\ee
where the sum is over the reduced Brillouin zone of Fig.~\ref{fig:four_Qs_BZ}(a), and we define $A_k\equiv\Phi_Q f_k+\Phi_Q^* f_{k+Q}^*.$
A $(\pi/2,0)$ CDW perturbation written in its 8-component extended Nambu basis
\begin{widetext}
\be
\psi_k^\dagger =
\begin{pmatrix}
c_{k\uparrow}^\dagger & c_{-k\downarrow} & c_{k+2Q\uparrow}^\dagger & c_{-k-2Q\downarrow} & c_{k+Q\uparrow}^\dagger &
c_{-k-Q\downarrow} & c_{k+3Q\uparrow}^\dagger & c_{-k-3Q\downarrow}
\end{pmatrix}
\label{eq:extended_Nambu_PIHALF}
\ee
is written as
\be
H_{CDW}^{(\pi/2,0)}=\sum_k^\prime \psi_k^\dagger H_k \psi_k
\ee
where $H_k$ is given by
\be
\begin{pmatrix}
& & & & A_k^* & 0 & A_{k+3Q} & 0 \\
& & & & 0 & -A_{-k-Q}^* & 0 & -A_{-k} \\
& & & & A_{k+Q} & 0 & A_{k+2Q}^* & 0 \\
& & & & 0 & -A_{-k-2Q} & 0 & -A_{-k-3Q}^* \\
A_k & 0 & A_{k+Q}^* & 0 & & & & \\
0 & -A_{-k-Q} & 0 & -A_{-k-2Q}^* & & & & \\
A_{k+3Q}^* & 0 & A_{k+2Q} & 0 & & & & \\
0 & -A_{-k}^* & 0 & -A_{-k-3Q} & & & &
\end{pmatrix}
\ee
\end{widetext}
\subsection{Pair density waves}
Scanning tunneling microscopy experiments  have revealed the presence of modulations in the local density of states in the vortex cores of the cuprate $\mathrm{Bi}_2 \mathrm{Sr}_2 \mathrm{Ca}_1 \mathrm{Cu}_2 \mathrm{O}_{8+\delta}$\cite{hof01,hof02,mce01,han01,mce02,koh01}, and in some instances, in the absence of magnetic field\cite{how01,how02}. More recent measurements, conducted in the absence of magnetic field, measured the spatial dependence of the superconducting gap\cite{sle01}. Their finding was that the superconducting order parameter is modulated, corresponding to superconducting pairs with a net center-of-mass momentum, which has become known as a pair density wave.

In addition to considering the effects of spatial modulations in the charge density, we therefore can also consider adding (to the $d$SC hamiltonian) a term corresponding to a modulation in the pair density. A pair density wave of wave vector $\vect{Q}$ is written as
\be
\label{eq:HPDW}
H_{PDW}=\sum_{\substack{k\\ \alpha\beta}} \Big(\Theta_Q g_k c_{k+Q\alpha}^\dagger c_{-k\beta}^\dagger+\Theta_Q^* g_k^* c_{-k\beta}c_{k+Q\alpha}\Big).
\ee
or, with the definition
$B_k~\equiv~\Theta_Q~(g_k+g_{-k-Q})$
we can write (for $(\pi,0)$ or $(\pi,\pi)$ density waves)
\bea
\label{eq:PDW}
H_{PDW}=\sum_k^\prime \psi_k^\dagger
\begin{pmatrix}
0 & 0 & 0 & B_k \\
0 & 0 & B_{-k}^* & 0 \\
0 & B_{-k} & 0 & 0 \\
B_k^* & 0 & 0 & 0
\end{pmatrix}
\psi_k
\eea
\subsection{Spin density waves}
The effective hamiltonian corresponding to a spin density wave of wave vector $\vect{Q}$ is
\bea
\label{eq:HSDW}
H_{SDW}=\sum_{k\sigma} \sigma \Big( \Phi_Q f_k c_{k+Q\sigma}^\dagger c_{k\sigma}+\Phi_Q^* f_k^* c_{k\sigma}^\dagger c_{k+Q\sigma} \Big).
\eea
In the extended Nambu basis, for a $\vect{Q}=(\pi,0)$ or $\vect{Q}=(\pi,\pi)$ SDW, this takes the form
\bea
H_{SDW}=\sum_k^\prime \psi_k^\dagger
\begin{pmatrix}
0 & 0 & A_k^* & 0 \\
0 & 0 & 0 & A_{-k} \\
A_k & 0 & 0 & 0 \\
0 & A_{-k}^* & 0 & 0
\end{pmatrix}
\psi_k,
\eea
where $A_k~\equiv~\Phi_Qf_k+\Phi_Q^*f_{k+Q}^*$.
\subsection{Checkerboard density waves}
In addition to broken symmetry states arising due to a single density wave, we can also consider multiple density waves. Scanning tunneling microscopy
experiments have previously revealed the presence of checkerboard order in $\textrm{BiSCCoO}$\cite{hof01,hof02,how01,ver01,mce01,han01}. While the wave
vectors of the order in those experiments was seen to be near $\vect{Q}\approx \frac{\pi}{2}$, for simplicity we first write down the hamiltonian
corresponding to $\vect{Q}_1=(\pi,0)$ and $\vect{Q}_2=(0,\pi)$ checkerboard order. The Brillouin zone is reduced to one fourth of its size, as is seen in
Fig.~\ref{fig:four_Qs_BZ}(c). The extended Nambu vector which describes such a system is
\begin{widetext}
\be
\psi_k^\dagger =
\begin{pmatrix}
c_{k\uparrow}^\dagger & c_{-k\downarrow} & c_{k+Q_x\uparrow}^\dagger & c_{-k-Q_x\downarrow} & c_{k+Q_y\uparrow}^\dagger & c_{-k-Q_y\downarrow} &
c_{k+Q_x+Q_y\uparrow}^\dagger & c_{-k-Q_x-Q_y\downarrow}
\end{pmatrix}
\ee
and the second quantized hamiltonian which describes the addition of a charge density wave and pair density wave is
\be
H_{\mathrm{CDW}}^{\mathrm{checkerboard}}+H_{\mathrm{PDW}}^\mathrm{checkerboard}=\sum^\prime_k \psi_k^\dagger H_k \psi_k,
\ee
where $H_k$ is given by
\be
\begin{pmatrix}
0 & 0 & A_k^{(x)*} & B_k^{(x)} & A_k^{(y)*} & B_k^{(y)} & 0 & 0 \\
0 & 0 & B_{-k}^{(x)*} & -A_{-k}^{(x)} & B_{-k}^{(y)*} & -A_{-k}^{(y)} & 0 & 0 \\
A_k^{(x)} & B_{-k}^{(x)} & 0 & 0 & 0 & 0 & A_{k+Q_x}^{(y)*} & B_{k+Q_x}^{(y)} \\
B_k^{(x)*} & -A_{-k}^{(x)*} & 0 & 0 & 0 & 0 & B_{-k-Q_x}^{(y)*} & -A_{-k-Q_x}^{(y)}\\
A_k^{(y)} & B_{-k}^{(y)} & 0 & 0 & 0 & 0 & A_{k+Q_y}^{(x)*} & B_{k+Q_y}^{(x)} \\
B_k^{(y)*} & -A_{-k}^{(y)} & 0 & 0 & 0 & 0 & B_{-k-Q_y}^{(x)} & -A_{-k-Q_y}^{(x)}\\
0 & 0 & A_{k+Q_x}^{(y)} & B_{-k-Q_x}^{(y)} & A_{k+Q_y}^{(x)} & B_{-k-Q_y}^{(x)} & 0 & 0 \\
0 & 0 & B_{k+Q_x}^{(y)*} & -A_{-k-Q_x}^{(y)*} & B_{k+Q_y}^{(x)*} & -A_{-k-Q_y}^{(x)*} & 0 & 0
\end{pmatrix}
\ee
\end{widetext}
and where
\bea
A_k^{(x)}\equiv \Phi_{Q_x}f_{k_x}+\Phi_{Q_x}^*f_{k+Q_x}^*\nonumber\\
A_k^{(y)}\equiv \Phi_{Q_y}f_{k_y}+\Phi_{Q_x}^*f_{k+Q_y}^*\nonumber\\
B_k^{(x)}\equiv \Theta_{Q_x}(g_{k_x}+g_{-k_x-Q_x})\nonumber\\
B_k^{(y)}\equiv \Theta_{Q_y}(g_{k_y}+g_{-k_y-Q_y})
\eea
represent the amplitudes of the charge density and pair density waves in the $x$ and $y$ directions.
\section{Thermal conductivity}
\label{sec:thermalcond}
At low temperatures, the temperature dependent phonon contribution to thermal conductivity vanishes as a power, $\kappa_{\mathrm{phonon}}\sim T^\alpha$.
\cite{chi01,chi02,sun01,sut01,haw01,tai01,pro01,haw02,hus01,and01,sun03,sun04} Therefore, the $T$-linear quasiparticle current can be extracted from experimental
data by plotting the measured thermal conductivity divided by temperature as a function of $T^{\alpha -1}$. In previous work\cite{dur02,sch01}, we considered a site-centered $\vect{Q}=(\pi,0)$ charge density wave and calculated the thermal conductivity using Green's functions obtained from the self-consistent Born approximation, and incorporated vertex corrections within the ladder approximation. Because the results of this work indicated that vertex corrections can usually be neglected, in what follows we will derive the thermal conductivity using the bare-bubble correlation function. This will greatly simplify the calculation, allowing its application to a variety of systems, whereby the thermal conductivity can be computed numerically given an an effective hamiltonian $H=H_{d\mathrm{SC}}+H_{\mathrm{DW}}(\psi)$ and an effective scattering rate $\Gamma_0$.
\subsection{Current operators}
In order to calculate the thermal conductivity, we first need to derive the heat-current associated with the quasiparticles. Because heat and spin currents are both
proportional to the quasiparticle current, we can get the heat-current by calculating the spin-current, and then using the energy measured from the Fermi
level as the associated charge (instead of the spin). To calculate the spin current for any particular hamiltonian, we write the density operator in second
quantized form, and then use Heisenberg equations of motion to find the momentum space representation of the current, that is
\bea
\lim_{q\rightarrow 0}(\vect{q}\cdot\vect{j^s})=[\rho_q^S,H].
\eea
The density operator is
\be
\rho_q^s=\sum_k^\prime \Big(c_{k\uparrow}^\dagger c_{k+q\uparrow} + c_{-k\downarrow}c_{-k-q\downarrow}^\dagger\Big).
\ee
Taking the commutator with the hamiltonians of Eqs. (\ref{eq:HdSC}),(\ref{eq:HCDW}), (\ref{eq:HPDW}) and (\ref{eq:HSDW}), using anti-commutation relations,
and discarding boundary terms, we find
\bea
[\rho_q^s,H]&=&\sum_{\substack{kk^\prime\\\sigma}}[\sigma c_{k^\prime\sigma}^\dagger c_{k^\prime+q\sigma},
\psi_k^\dagger \wt{H}_k \psi_k]\nonumber\\
&=&\sum_k \vec{q}\cdot \psi_k^\dagger \frac{\partial \wt{H}_k}{\partial \vec{k}} \psi_k
\eea
for the spin current. The heat current in the Matsubara representation is given by
\be
\wt{j}(i\omega,i\Omega)=(i\omega+\frac{i\Omega}{2})\sum_k\psi_k^\dagger\frac{\partial \wt{H}_k}{\partial \vec{k}}\psi_k
\ee
Now we have a generalized velocity operator in the Nambu space, $\wt{v}_k = \frac{\partial \wt{H}_k}{\partial\vec{k}}$. For instance, for the $\vec{Q}=(\pi,0)$ pair density wave of Eq. \ref{eq:PDW}, the velocity operator would read
\be
\wt{v}(\vec{k})=
\begin{pmatrix}
\vec{v}_{f,k} & \vec{v}_{\Delta,k} & 0 & \frac{\partial A_k}{\partial\vec{k}} \\
\vec{v}_{\Delta,k} & -\vec{v}_{f,k} & \frac{\partial A_{-k}^*}{\partial\vec{k}} & 0 \\
0 & \frac{\partial A_{-k}}{\partial\vec{k}} & \vec{v}_{f,k+Q} & \vec{v}_{\Delta,k+Q} \\
\frac{\partial A_k^*}{\partial\vec{k}} & 0 & \vec{v}_{\Delta,k+Q} & -\vec{v}_{f,k+Q}
\end{pmatrix}
\ee
For density waves without internal momentum dependance, or for those where the variation is slight near the nodal locations, the velocity operator reduces
to the form found in Refs.~\onlinecite{dur02,sch01}
\be
\wt{v}_{f,k}=
\begin{pmatrix}
\vec{v}_{f,k} & \vec{v}_{\Delta,k} \\
\vec{v}_{\Delta,k} & -\vec{v}_{f,k}
\end{pmatrix}
\ee
\subsection{Universal limit thermal conductivity}
The universal limit thermal conductivity is calculated using linear response formalism. The
thermal conductivity is given in terms of the retarded current-current correlation function.
\be
\frac{K(\Omega,T)}{T}=\lim_{\Omega\rightarrow 0}-\frac{\textrm{Im}(\Pi_{\textrm{Ret}}(\Omega))}{\Omega T^2}
\ee
We evaluate the correlation function using the Matsubara method\cite{mahan}. The bare-bubble correlator, given in terms of a spectral representation, is
\be
\label{eq:current-current}
\Pi(i\Omega)=\int \mathrm{d}\omega_1 \mathrm{d}\omega_2 \mathrm{Tr} \sum_k \Big[\wt{A}(\omega_1)\wt{v}
\wt{A}(\omega_2)\wt{v}\Big] S(i\Omega)
\ee
where
\bea
S(i\Omega)\equiv \sum_{i\omega_n}\frac{(i\omega+\frac{i\Omega}{2})^2}{(i\omega-\omega_1)(i\omega+i\Omega-\omega_2)}
\eea
and $A(\vec{k},\omega)$ is the spectral function.

It is important to use the correct form of the spectral function in Eq.~(\ref{eq:current-current}) to avoid erroneous results, as is noted in Ref.~\onlinecite{dur02}.
For example, a bond-centered CDW of wave vector $\vec{Q}=(\pi,0)$, which looks like
\be
H_{CDW}=
\begin{pmatrix}
0 & 0 & i\psi & 0\\
0 & 0 & 0 & -i\psi \\
i\psi & 0 & 0 & 0 \\
0 & -i\psi & 0 & 0
\end{pmatrix}
\ee
leads to a spectral function which is not real, and the spectral function is not given by the formula
\be
\wt{A}(\vec{k},\omega)=-\frac{1}{\pi}\textrm{Im}(G_R(\vec{k},\omega)),
\ee
but rather by
\be
A(\vec{k},\omega)\equiv \frac{-1}{2\pi i}\Big(G_R(\vec{k},\omega)-G_A(\vec{k},\omega)\Big).
\label{eq:spectral}
\ee
The details of the thermal conductivity calculation are similar to those of Refs.~\onlinecite{dur02,sch01}. In general, the self-consistent t-matrix approximation
can be used to compute the Green's functions, however, here we use a simpler, diagonal self-energy as a first approximation. In terms of a model hamiltonian
$H_k$, and incorporating impurity scattering by assuming an imaginary part of the self-energy, $\wt{\Sigma}_R(\omega\rightarrow 0)=-i\Gamma_0$, the
universal limit thermal conductivity is
\be
\lim_{T\rightarrow 0}\frac{\kappa_0}{T}=\frac{k_B^2 \pi^2}{3}\sum_k \mathrm{Re}\Big[\mathrm{Tr}[\wt{A}(0)\frac{\partial \wt{H}_k}{\partial \vec{k}}\wt{A}(0)
\frac{\partial\wt{H}_k}{\partial \vec{k}}]\Big]
\label{eq:kappa}
\ee
where
\bea
\wt{G}_R(\vect{k},\omega)=\Big(\omega-\wt{H}_k+i\Gamma(\omega)\Big)^{-1}\nonumber\\
\wt{G}_A(\vect{k},\omega)=\Big(\omega-\wt{H}_k-i\Gamma(\omega)\Big)^{-1}.
\eea
\section{Effects on energy spectrum and thermal conductivity}
\label{sec:results}
Here we modify the $d$SC hamiltonian of Eq.~(\ref{eq:HdSC}) with the addition of density waves such as those of Eqs.~(\ref{eq:HCDW}), (\ref{eq:HPDW}) and (\ref{eq:HSDW}), which will be tuned
by the real parameter $\psi$, the strength of the density wave. This is done to study the behavior of the quasiparticle spectrum, and through Eq.~(\ref{eq:kappa}),
the universal limit thermal conductivity. In each of the figures from Fig.~\ref{fig:PIZEROCDW} to Fig.~\ref{fig:Checkerboard_PDW}, we present (a) The trajectory of the nodes in the region $0<k_x,k_y<\frac{\pi}{2}$, as the density wave is turned on (the starting place (node for $d$SC system) is indicated with a star), (b) (minimum) quasiparticle energy as a function of the order parameter strength $\psi$, and (c) universal limit thermal conductivity as a function of $\psi$. In all instances, the universal limit conductivity $\frac{\kappa_{00}}{T}$ is given in units of $\frac{k_B^2}{3\hbar}\frac{v_f^2+v_\Delta^2}{v_f v_\Delta}$, the value for the original $d$SC system, and we measure $\Delta_0$, $E_{\mathrm{min}}$, $\mu$ and $\Gamma_0$ in units of $t$, the hopping parameter.
\subsection{$\vect{Q}=(\pi,0)\,\,\,$ density waves}
\begin{figure}[htbp]
  	\centerline{
 \resizebox{3.25 in}{!}
  {\includegraphics{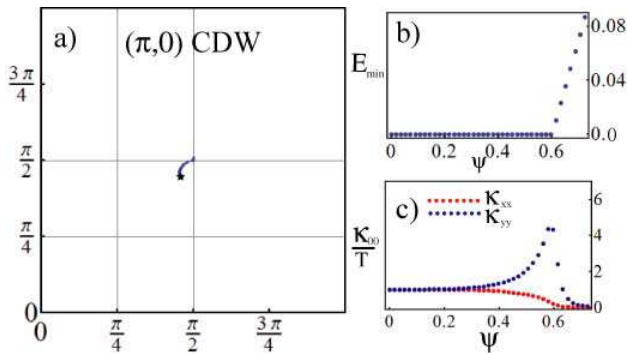}}}
  \caption[Effects of $\vect{Q}=(\pi,0)$ charge density wave]{Effects on spectrum and low temperature transport of a $\vect{Q}=(\pi,0)$ charge density wave. The results are the same for site centered ($s$-wave or $p_y$-wave) and bond centered ($p_x$-wave) density waves, in that the nodes evolve along the same curved paths toward the $(\pm\pi/2,\pm\pi/2)$
  points where they collide with their images from the next reduced Brillouin zone. As this happens, the nodes are nested and the spectrum is gapped. The universal
  limit thermal conductivity vanishes beyond this point. Disorder $\Gamma_0$ broadens the transition. Here we take $\mu=-0.6$, $\Delta_0=4$ and $\Gamma_0=0.02$.}
  \label{fig:PIZEROCDW}
\end{figure}

\begin{figure}[htbp]
  	\centerline{
 \resizebox{3.25 in}{!}
  {\includegraphics{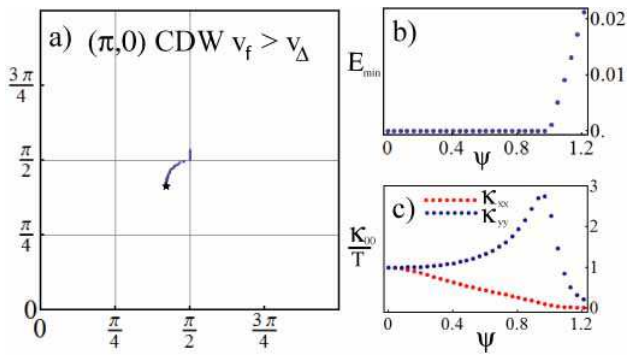}}}
  \caption[Effects of $\vect{Q}=(\pi,0)$ charge density wave, $v_f>v_\Delta$]{Effects on spectrum and low temperature transport of a $\vect{Q}=(\pi,0)$ charge density wave. Depicted are the results for $\mu=-1$, $\Delta_0=0.4$
   and $\Gamma_0 =0.02$. These parameters describe anisotropic Dirac quasiparticles, with $v_f/v_\Delta=10$. The anisotropy tends to suppress $\kappa_{00}$ slightly.}
  \label{fig:PIZEROCDW_anisotropic}
\end{figure}

The addition of a $\vect{Q}=(\pi,0)$ charge density wave to a $d$-wave superconductor has been considered before\cite{par01,dur02,sch01}. As the perturbation
is turned on, the nodes' locations evolve along curved paths, until they meet the images of the nodes from the second reduced Brillouin zone at the collision
point $(\pi/2,\pi/2)$, as seen in Figs.~\ref{fig:PIZEROCDW} and \ref{fig:PIZEROCDW_anisotropic} . The effect is the same, regardless of whether the density wave is of $s$-wave $(\Phi_Q=\psi, f_k=1$, site-centered), $p_x$-wave $(\Phi_Q=i\psi, f_k=\sin(k_x a)$, bond-centered) or $p_y$-wave $(\Phi_Q=\psi, f_k=\sin(k_y a)$, site-centered) symmetry. The critical value of $\psi$ which gaps the system is $\psi_c=v_f k_0$, where
\be
k_0=\sqrt{2}\Big[\frac{\pi}{2}-\cos^{-1}\Big(\frac{-t}{2t^\prime}+\sqrt{(\frac{t}{2t^\prime})^2-\frac{\mu}{4t}}\Big)\Big]
\ee
 is the distance separating the $\psi=0$ nodal point from $(\pi/2,\pi/2)$ in $k$-space. For an $s$-wave perturbation of strength $\psi$ representing a charge, pair, or spin density wave, the quasiparticle spectrum is
 \bea
 \omega=\sqrt{A-\sqrt{A^2+B-C}},
 \eea
 where
 \bea
 A&\equiv &\frac{\epsilon_k^2+\Delta_k^2+\epsilon_{k+Q}^2+\Delta_{k+Q}^2}{2}+\psi^2\nonumber\\
 B&\equiv &2\psi^2(a\epsilon_k\epsilon_{k+Q}+ b\Delta_k \Delta_{k+Q})\nonumber\\
 C&\equiv &(\epsilon_k^2+\Delta_k^2)(\epsilon_{k+Q}^2+\Delta_{k+Q}^2),
 \eea
 and $(a,b)=(-1,1)$ for a charge density wave, $(1,-1)$ for a pair density wave, and $(-1,-1)$ for a spin density wave.

\begin{figure}[htbp]
  	\centerline{
 \resizebox{3.25 in}{!}
  {\includegraphics{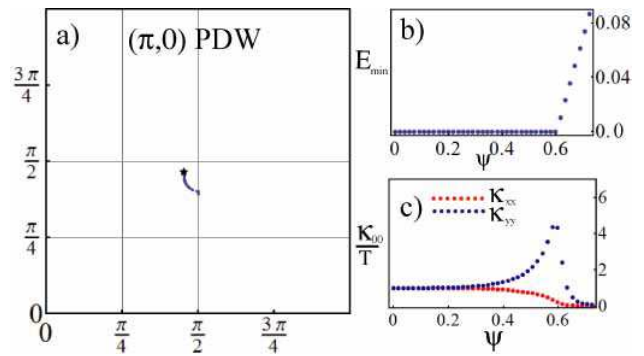}}}
  \caption[Effects of $\vect{Q}=(\pi,0)$ pair density wave]{Effects on spectrum and low temperature transport of a $\vect{Q}=(\pi,0)$ pair density wave. Depicted are the results for $\mu=-0.6$, $\Delta_0=4$
  ($v_f=v_\Delta$) and $\Gamma_0=0.02$. As was the case for the CDW, the nodes evolve along a curved path towards the $(\pi/2,k_y)$ line. Upon reaching $k_x=\pi/2$,
   the nodes are nested, and the spectrum is gapped. For $\psi$ larger than the critical value $\psi_c$, the thermal conductivity vanishes, up to disorder broadening.}
    \label{fig:PIZEROPDW}
\end{figure}

 The resulting thermal conductivity is anisotropic, reflecting the striped nature of the system. The nodes are deformed as they approach the collision point, and the thermal conductivity
$\kappa_{yy}$ perpendicular to the direction of the density wave increases at first, before both $\kappa_{xx}$ and $\kappa_{yy}$ vanish for larger amplitudes of density wave,
 $\psi$. The effect of a $\vect{Q}=(\pi,0)$ pair density wave is similar to that of the charge density wave: the nodes evolve along a curved path until they meet
  their images in the second reduced Brillouin zone, and the resulting universal limit thermal conductivity is the same. The effects of a site-centered $(\pi,0)$
  pair density wave is shown in Fig.~\ref{fig:PIZEROPDW}.

\begin{figure}[htbp]
  	\centerline{
 \resizebox{3.25 in}{!}
  {\includegraphics{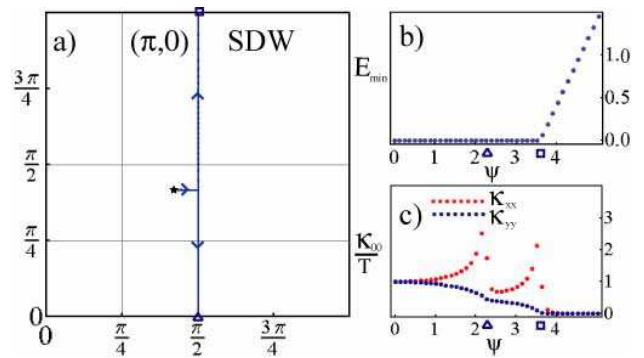}}}
  \caption[Effects of $\vect{Q}=(\pi,0)$ spin density wave]{Effects on spectrum and low temperature transport of a $\vect{Q}=(\pi,0)$ spin density wave. Depicted are the results for parameters $\Delta_0=4$, $\mu=-0.6$ and $\Gamma_0=0.02$.
  As the density wave is turned on, the nodes move in a straight line to the $(\pm\pi/2,k_y)$ lines. When they reach that line, each node splits in two, and the two nodes move up and down along that line. The spectrum remains gapless, even though the nodes are nested by the ordering vector. Correspondingly, the thermal conductivity is unaffected at that energy scale. For $\psi$ much larger, these two nodes collide with their images in the second reduced Brillouin zone (at different values of $\psi$), and the thermal conductivity is reduced by one half of the pure $d$SC value after each such collision. The locations of the two separate nodal collisions are illustrated by a square and a triangle in (a), and the order strength at which they appear is given in (b) and (c).}
  \label{fig:PIZEROSDW}
\end{figure}

A more unusual case is that of the $\vect{Q}=(\pi,0)$ spin density wave. With this perturbation, the nodal points evolve directly towards the $(\pi/2,k_y)$ line, as seen in Fig.\ref{fig:PIZEROSDW}. The quasiparticle spectrum then evolves so that there are two minima. In other words, the node splits in two, and nodes move up and down the $(\pi/2,k_y)$ line. The nodes are nested by $\vect{Q}$, but the spectrum remains gapless, and the universal limit thermal conductivity is unaffected. If the perturbation is allowed to become extremely large ($\psi>>\Delta_0$), then the nodes (there are now twice as many, each on a reduced Brillouin zone edge) collide with their images from the second reduced Brillouin zone, and the thermal conductivity then vanishes. The split-off nodes collide at different strengths of $\psi$, however, and the spectral weight disappears in two steps, accordingly, as does the thermal conductivity. The fact that the nodal structure is preserved in this case, even when the nodes become nested, runs contrary to the intuition (suggested by the converse of the theorem of Ref.\onlinecite{ber01})  that such nested nodes would become gapped.
\subsection{$\vect{Q}=(\pi,\pi)\,\,\,$ density waves}
\begin{figure}[htbp]
  	\centerline{
 \resizebox{3.25 in}{!}
  {\includegraphics{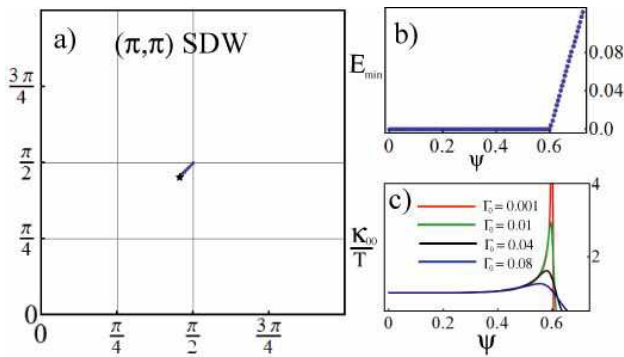}}}
  \caption[Effects of $\vect{Q}=(\pi,\pi)$ charge density wave]{Effects on spectrum and low temperature transport of a $\vect{Q}=(\pi,\pi)$ spin density wave.
  As the density wave is turned on, the nodes move along the symmetry lines $k_x=\pm k_y$ toward $(\pm \pi/2,\pm \pi/2)$, where they become gapped. Accordingly,
  $\kappa_{00}$ vanishes. In (c) the effects of increasing disorder are presented. The disorder tends to smear the thermal conductivity around the nodal transition;
  as such, $\kappa_{00}$ is no longer universal.
  }
  \label{fig:PIPISDW}
\end{figure}

Adding a $\vect{Q}=(\pi,\pi)$ spin density wave was also discussed as an example in Ref.~\onlinecite{ber01}. In real space, such a density wave is modulated
as $\cos(k_x)\cos(k_y)$, so that nodes remain along the $(\pi,\pi)$ direction as the density wave is turned on, as is seen in Fig.~\ref{fig:PIPISDW}.
 When the nodes reach $(\pi/2,\pi/2)$, the system is gapped\cite{ber01}, and the thermal conductivity vanishes. On the other hand, a $\vect{Q}=(\pi,\pi)$ charge density wave behaves in a similar manner
to the $(\pi,0)$ spin density wave, in that the nodes do not vanish for small perturbations.

The addition of a $\vect{Q}=(\pi,\pi)$ pair density wave drives the location of the nodes towards the $\Gamma$ point at $(k_x,k_y)=(0,0)$, an effect which is also observed via the addition of the checkerboard pair density wave. In both instances, a large perturbation $\psi>>\Delta_0$ is required to affect the thermal conductivity.
\subsection{$\vect{Q}=(\pi/2,0)\,\,\,$ charge density wave}
\begin{figure}[htbp]
  	\centerline{
 \resizebox{3.25 in}{!}
  {\includegraphics{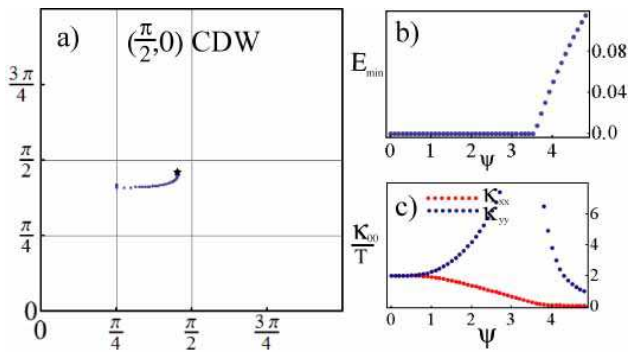}}}
  \caption[Effects of $\vect{Q}=(\pi/2,0)$ charge density wave]{Effects on spectrum and low temperature transport of a $\vect{Q}=(\pi/2,0)$ charge density wave. As the density wave is turned on, the nodes move in a
  curved path to the $(\pm\pi/4,k_y)$ lines. The spectrum becomes gapped at that point, when the node is nested by the ordering vector and the thermal conductivity
  vanishes for $\psi$ larger than about $4t$. Such a system is out of the range of validity of our model, as it would be dominated by the charge order, rather than
  the $d$-wave superconductor. For $\psi$ on the order of $\psi_c$, $\kappa_{00}$ retains its $d$SC value. Here, $\mu=-0.6$, $\Delta_0=4$ and $\Gamma_0=0.05$
  }
  \label{fig:PIHALF_CDW}
\end{figure}

A $\vect{Q}=(\pi/2,0)$ density wave behaves slightly differently from the $(\pi,0)$ case. In this case, the nodes are driven towards the $(\pi/4,k_y)$ line,
rather than $(\pi/2,k_y)$. While they would become gapped if they arrived there, for realistic parameters $t,\mu,$ and $\Delta_0$, such a density wave would
dominate the system, that is, $\psi>>\Delta_0$. The evolution is as seen in Fig.~\ref{fig:PIHALF_CDW}
and preserves the nodes for $\psi <\Delta_0$. As such, the universal limit thermal conductivity is not significantly affected by this perturbation.
\subsection{$\vect{Q}_1=(\pi,0)$, $\vect{Q}_2=(0,\pi)$ checkerboard density waves}
\begin{figure}[htbp]
  	\centerline{
 \resizebox{3.25 in}{!}
  {\includegraphics{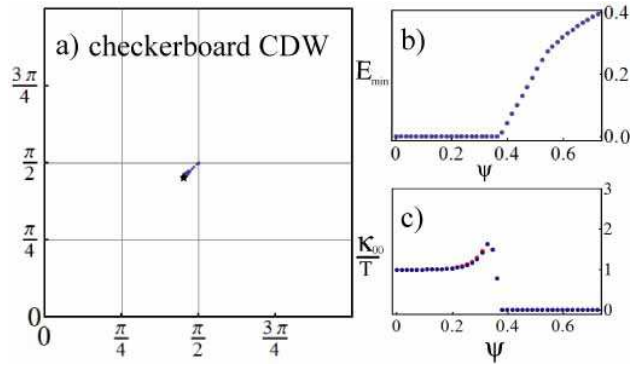}}}
  \caption[Effects of $\vect{Q}_1=(\pi,0)$, $\vect{Q}_2=(0,\pi)$ checkerboard charge density wave]{Effects on spectrum and low temperature transport of a $\vect{Q}_1=(\pi,0)$, $\vect{Q}_2=(0,\pi)$ charge density perturbation to the $d$SC system.
  The nodes move in straight lines toward the $(\pm\pi/2,\pm\pi/2)$ points. The spectrum becomes gapped at that point, and $\kappa_{00}$ vanishes for $\psi$
  larger than about $0.4 t$.}
  \label{fig:checkerboard_CDW}
\end{figure}

\begin{figure}[htbp]
  	\centerline{
 \resizebox{3.25 in}{!}
  {\includegraphics{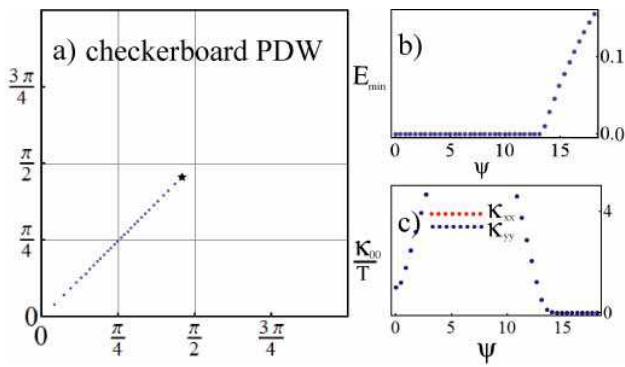}}}
  \caption[Effects of $\vect{Q}_1=(\pi,0)$, $\vect{Q}_2=(0,\pi)$ checkerboard pair density wave]{Effects on spectrum and low temperature transport of a $\vect{Q}_1=(\pi,0)$, $\vect{Q}_2=(0,\pi)$ pair density perturbation to the $d$SC system.
  The nodes move in straight lines toward the origin. The spectrum would become gapped at that point, however, the value of $\psi$ required is
  much larger than the energy scale of the superconducting order parameter. Therefore, for reasonable strengths of the ordering vector, the thermal
  conductivity is unaffected by this density wave.}
  \label{fig:checkerboard_PDW}
\end{figure}

Configurations with more than one density wave can also be considered in this formalism. In this paper, we turn our attention to the checkerboard configuration
illustrated in part (c) of Fig.~\ref{fig:four_Qs_BZ}. As we turn on two charge density waves of $\vect{Q}_1=(\pi,0)$ and $\vect{Q}_2=(0,\pi)$,
with equal amplitudes, the nodes are perturbed along the symmetry line toward the $(\pi/2,\pi/2)$ point, as shown in Fig.~\ref{fig:checkerboard_CDW}.
When the nodes reach the $(\pi/2,\pi/2)$ point, the spectrum becomes gapped, and the thermal conductivity vanishes, with a value of $\psi_c$ about two-thirds
of that for the striped $(\pi,0)$ CDW. In contrast, the checkerboard pair density wave seen in Fig.~\ref{fig:checkerboard_PDW} evolves the nodes along the
same symmetry line, but towards the $\Gamma$ point $(0,0)$. At that point, the spectrum would become gapped, and the universal limit thermal conductivity would
vanish. However, systems which more closely resemble a $d$-wave superconductor than the checkerboard $(\Delta_0 > \psi)$ will remain gapless, as the nodal
evolution would not be driven that far---about thirty times the critical value for the striped $(\pi,0)$ PDW .
\section{Conclusions}
In conclusion, we have written mean-field hamiltonians describing a $d$-wave superconductor perturbed by a variety of density waves. We noted the effects
of such perturbations on the low energy quasiparticle spectrum, and by calculating the universal limit $(T\rightarrow 0,\Omega\rightarrow 0)$ thermal
conductivity, see the effect that density waves can have on the low temperature thermal transport. Whether or not the universal limit thermal
conductivity is robust in the presence of an incipient density wave depends on which type of density wave, and which wave vector, is added. For instance, in
the case of $\vect{Q}=(\pi,0)$ pair density waves, the quasiparticle nodes evolve so that their $k$-space locations move toward $(\pm\pi/2,\pm\pi/2)$,
as they do for a CDW of the same wave vector. When they reach this point, which is the point at which the density wave vector nests the nodes, the spectrum becomes
gapped. However, for the $\vect{Q}=(\pi,0)$ spin density wave, the nodal structure is preserved beyond this ordering strength, despite the nesting of the nodes. For the $\vec{Q}=(\pi,\pi)$ density waves, the effects of SDW and CDW are reversed from that of the $(\pi,0)$ case; the $(\pi,\pi)$ CDW preserves nodality beyond the nesting wave vector, while the $(\pi,\pi)$ SDW is gapped beyond a critical strength. In the case of $\vect{Q}=(\pi/2,0)$, the different wave vector drives the nodes toward $(\pi/4,k_y)$ instead. Given typical tight-binding parameters, such a charge density wave will not gap the quasiparticle spectrum, and will thus not affect the thermal conductivity, which remains universal. In the case of $\vect{Q}_1=(\pi,0)$, $\vect{Q}_2=(0,\pi)$ checkerboard charge order, whether or not the universal limit thermal conductivity is robust depends on which type of density wave is present. The CDW checkerboard nodes move toward the $(\pm\pi/2,\pm\pi/2)$ point, and become gapped. However, the PDW checkerboard nodes move away from that direction and the nodal structure is preserved.

Because the onset of charge ordering is believed to be correlated with underdoping, observations which show that the low temperature thermal conductivity differs from the universal value predicted in Ref.~\onlinecite{dur01} may be due to the influence of coexisting orders.
There are some general features that appear in all of the models considered in this paper.
(1) In general, the nodal evolution is determined more by the wave-vector $\vect{Q}$ than by the chemical potential $\mu$, although $\mu$ will determine the amplitude of density wave which will gap the system.
(2) The physics still remains nodal in the following sense. Whether the density waves considered were of $s$-wave or $p$-wave symmetry did not have an effect; all that matters is the amplitude of the density wave at the node.
(3) It is interesting to note that the universal limit thermal conductivity generally develops a disorder dependence, especially near the nodal transition point. The presence of density waves are therefore one possible explanation of the breakdown of universal limit thermal transport in cuprates.
(4) In general, there is an increase in the thermal conductivity near the nodal transitions (for cases where there are), which is caused by the deformation of the nodes (and the resulting effective change in $v_f/v_\Delta$) as they meet their images in the second reduced Brillouin zone. This feature is consistent with the thermal conductivity measurements of Proust {\em et al}.\cite{pro02} who find a large enhancement in thermal conductivity of Bi-2201.

\acknowledgments{We are grateful to Subir Sachdev and Steve Kivelson for helpful discussions. This work was supported by NSF Grant No. DMR-0605919.}



\begin{thebibliography}{50}
\bibitem{lee01} P.A. Lee, Phys.\ Rev.\ Lett.\ {\bf 71}, 1887 (1993). 
\bibitem{dur01} A.C. Durst and P.A. Lee, Phys.\ Rev.\ B {\bf 62} 1270 (2000). 
\bibitem{gra01} M.J. Graf, S-K. Yip, J.A. Sauls and D. Rainer, Phys.\ Rev.\ B {\bf 53} 15147 (1996).
\bibitem{tai01} L. Taillefer, B. Lussier, R. Gagnon, K. Behnia and H. Aubin, Phys.\ Rev.\ Lett.\ {\bf 79}, 483 (1997).
\bibitem{chi01} M. Chiao, R.W. Hill, C. Lupien, B. Popi$\acute{c}$, R. Gagnon and L. Taillefer, Phys.\ Rev.\ Lett.\ {\bf 82} 2943 (1999)
\bibitem{chi02} M. Chiao, R.W. Hill, C. Lupien, L. Taillefer, P. Lambert, R. Gagnon and P. Fournier, Phys.\ Rev.\ B {\bf 62} 3554 (2000)
\bibitem{nak01} S. Nakamae, K. Behnia, L. Balicas, F. Rullier-Albenque, H. Berger and T. Tamegai, Phys. Rev. B {\bf 63} 184509 (2001) 
\bibitem{pro01} C. Proust, E. Boaknin, R.W. Hill, L. Taillefer and A.P. Mackenzie, Phys.\ Rev.\ Lett.\ {\bf 89}, 147003 (2002).
\bibitem{sut01} M. Sutherland, D.G. Hawthorn, R.W. Hill, F. Ronning, S. Wakimoto, H. Zhang, C. Proust, E. Boaknin, C. Lupien, L. Taillefer, R.X. Liang, D.A. Bonn,
                 W.N. Hardy, R. Gagnon, N.E. Hussey, T. Kimura, M. Nohara and H. Takagi, Phys. Rev. B {\bf 67} 174520 (2003)
\bibitem{hil01} R.W. Hill, C. Lupien, M. Sutherland, E. Boaknin, D.G. Hawthorn, C. Proust, F. Ronning, L. Taillefer, R. Liang, D.A. Bonn and W.N. Hardy, Phys.
                 \ Rev.\ Lett.\ {\bf 92}, 027001 (2004).
\bibitem{sun01} X.F. Sun, K. Segawa and Y. Ando, Phys.\ Rev.\ Lett.\ {\bf 93}, 107001 (2004).
\bibitem{sut02} M. Sutherland, S.Y. Li, D.G. Hawthorn, R.W. Hill, F. Ronning, M.A. Tanatar, J. Paglione, H. Zhang, L. Taillefer, J. DeBenedictis, R. Liang,
                D.A. Bonn and W.N. Hardy, Phys.\ Rev.\ Lett.\ {\bf 94}, 147004 (2005).
\bibitem{haw01} D.G. Hawthorn, S.Y. Li, M. Sutherland, E. Boaknin, R.W. Hill, C. Proust, F. Ronning, M.A. Tanatar, J.P. Paglione, L. Taillefer, D. Peets,
                R.X. Liang, D.A. Bonn, W.N. Hardy and N.N. Kolesnikov, Phys. Rev. B {\bf 75} 104518 (2007) 
\bibitem{sun02} X.F. Sun, S. Ono, X. Zhao, Z.Q. Pang, Y. Abe and Y. Ando, Phys.\ Rev.\ B {\bf 77} 094515 (2008).
\bibitem{hus01} N.E. Hussey, Advances in Physics, {\bf 51}, 1685 (2002).
\bibitem{haw02} D.G. Hawthorn, R.W. Hill, C. Proust, F. Ronning, M. Sutherland, E. Boaknin, C. Lupien, M.A. Tanatar, J. Paglione, S. Wakimoto, H. Zhang,
                L. Taillefer, T. Kimura, M. Nohara, H. Takagi and N.E. Hussey, Phys.\ Rev.\ Lett.\ {\bf 90}, 197004 (2003).
\bibitem{and01} Y. Ando, S. Ono, X.F. Sun, J. Takeya, F.F. Balakirev, J.B. Betts and G.S. Boebinger, Phys.\ Rev.\ Lett.\ {\bf 92}, 247004 (2004).
\bibitem{sun03} X.F. Sun, K. Segawa and Y. Ando, Phys.\ Rev.\ B {\bf 72}, 100502(R) (2005).
\bibitem{sun04} X.F. Sun, S. Ono, Y. Abe, S. Komiya, K. Segawa and Y. Ando, Phys.\ Rev.\ Lett.\ {\bf 96}, 017008 (2006).
\bibitem{ander01} B.M. Andersen and P.J. Hirschfeld, Phys. Rev. Lett. {\bf 100} 257003 (2008).
\bibitem{pro02} C. Proust, K. Behnia, R. Bel, D. Maude and S.I. Vedeneev, Phys. Rev. B {\bf 72} 214511 (2005).
\bibitem{hof01} J.E. Hoffmann, E.W. Hudson, K.M. Lang, V. Madhavan, H. Eisaki, S. Uchida, and J.C. Davis, Science {\bf 295} 466 (2002) 
\bibitem{hof02} J.E. Hoffmann, K. McElroy, D.H. Lee, K.M. Lang, H. Eisaki, S. Uchida and J.C. Davis, Science {\bf 297}, 1148 (2002).
\bibitem{how01} C. Howald, H. Eisaki, N. Kaneko, M. Greven and A. Kapitulnik, Phys. Rev. B {\bf 67} 014533 (2003) 
\bibitem{ver01} M. Vershinin, S. Misra, S. Ono, Y. Abe, Y. Ando and A. Yazdani, Science {\bf 303}, 1995 (2004).
\bibitem{mce01} K. McElroy, D.H. Lee, J.E. Hoffman, K.M. Lang, J. Lee, E.W. Hudson, H. Eisaki, S. Uchida, and J.C. Davis, Phys.\ Rev.\ Lett.\ {\bf 94}, 197005 (2005) 
\bibitem{han01} T. Hanaguri, C. Lupien, Y. Kohsaka, D.H. Lee, M. Azuma, M. Takano, H. Takagi and J.C. Davis, Nature {\bf 430} 1001 (2004) 
\bibitem{mis01} S. Misra, M. Vershinin, P. Phillips and A. Yazdani, Phys.\ Rev.\ B {\bf 70} 220503(R) (2004).
\bibitem{mce02} K. McElroy, D.H. Lee, J.E. Hoffman, K.M. Lang, J. Lee, E.W. Hudson, H. Eisaki, S. Uchida and J.C. Davis, Phys.\ Rev.\ Lett.\ {\bf 94}, 197005 (2005).
\bibitem{koh01} Y. Kohsaka, C. Taylor, K. Fujita, A. Schmidt, C. Lupien, T. Hanaguri, M. Azuma, M. Takano, H. Eisaki, H. Takagi, S. Uchida and J.C. Davis, Science {\bf 315}, 1380 (2007).
\bibitem{boy01} M.C. Boyer, W.D. Wise, K. Chatterjee, M. Yi, T. Kondo, T. Takeuchi, H. Ikuta and E.W. Hudson, Nature Physics {\bf 3}, 802 (2007).
\bibitem{han02} T. Hanguri, Y. Kohsaka, J.C. Davis, C. Lupien, I. Yamada, M. Azuma, M. Takano, K. Ohishi, M. Ono and H. Takagi, Nature Physics {\bf 3}, 865 (2007).
\bibitem{pas01} A.N. Pasupathy, A. Pushp, K.K. Gomes, C.V. Parker, J. Wen, Z. Xu, G. Gu, S. Ono, Y. Ando and A. Yazdani, Science {\bf 320}, 196 (2008).
\bibitem{wis01} W.D. Wise, M.C. Boyer, K. Chatterjee, T. Kondo, T. Takeuchi, H. Ikuta, Y. Wang and E.W. Hudson, Nature Physics {\bf 4}, 696 (2008).
\bibitem{koh02} Y. Kohsaka, C. Taylor, P. Wahi, A. Schmidt, J. Lee, K. Fujita, J.W. Allredge, K. McElroy, J. Lee, H. Eisaki, S. Uchida, D.H. Lee and J.C. Davis, Nature {\bf 454}, 1072 (2008).
\bibitem{kiv01} S.A. Kivelson, I.P. Bindloss, E. Fradkin, V. Oganesyan, J.M. Tranquada, A. Kapitulnik and C. Howald, Rev.\ Mod.\ Phys.\ {\bf 75}, 1201 (2003) (and references within).
\bibitem{fis01} {\O}. Fischer, M. Kugler, I. Maggio-Aprile and C. Berthod, Rev.\ Mod.\ Phys.\ {\bf 79} 353 (2007).
\bibitem{cha01} S. Chakravarty, R.B. Laughlin, D.K. Morr and C. Nayak Phys.\ Rev.\ B {\bf 63} 094503 (2001).
\bibitem{ber01} E. Berg, C.C. Chen and S.A. Kivelson, Phys.\ Rev.\ Lett.\ {\bf 100} 027003 (2008) 
\bibitem{par01} K. Park and S. Sachdev, Phys\. Rev\. B {\bf 64} 184510 (2001)
\bibitem{dur02} A.C. Durst and S. Sachdev, http://arxiv.org/abs/0810.3914v1 (2008) 
\bibitem{sch01} P.R. Schiff and A.C. Durst, Physica C {\bf 469} 740 (2009) 
\bibitem{gus01} V.P. Gusynin and V.A. Miransky, Eur. Phys. J. B 37, 363–368 (2004).
\bibitem{schrieffer01} R. Schrieffer, {\it Superconductivity}, (W.A. Benjamin Publishers, New York, 1964).
\bibitem{nay01} C. Nayak, Phys.\ Rev.\ B {\bf 62} 4880 (2000). 
\bibitem{how02} C. Howald, P. Fournier and A. Kapitulnik, Phys. Rev. B {\bf 64} 100504(R) (2001) 
\bibitem{sle01} J.A. Slezak, J.H. Lee, M. Wang, K. McElroy, K. Fujita, B.M. Andersen, P.J. Hirschfeld, H. Eisaki, S. Uchida and J.C. Davis, Proc. Nat. Acad. of Science, {\bf 105} 3203 (2008). 
\bibitem{mahan} G. Mahan, {\it Many-Particle Physics}, (Plenum Press, New York, 1981).
\end{thebibliography}
\end{document}